\documentclass[superscriptaddress,reprint,showpacs,amssymb,amsmath,aps,prl]{revtex4-1}
\usepackage{graphicx}
\usepackage{bm}
\usepackage{amsthm}
\usepackage{amsfonts}

\begin{document}
\title{Factorization and criticality in finite $XXZ$ systems of arbitrary spin}
\author{M. Cerezo}
\affiliation{Instituto de F\'{\i}sica de La Plata, CONICET, and Departamento de F\'{\i}sica,
	Universidad Nacional de La Plata, C.C.\ 67, La Plata 1900, Argentina}
\author{R. Rossignoli}
\affiliation{Instituto de F\'{\i}sica de La Plata, CONICET, and Departamento de F\'{\i}sica,
	Universidad Nacional de La Plata, C.C.\ 67, La Plata 1900, Argentina}
\affiliation{Comisi\'on de Investigaciones Cient\'{\i}ficas de la Provincia de Buenos Aires (CIC), La Plata (1900), Argentina}
\author{N. Canosa}
\affiliation{Instituto de F\'{\i}sica de La Plata, CONICET, and Departamento de F\'{\i}sica,
	Universidad Nacional de La Plata, C.C.67, La Plata 1900, Argentina}
\author{ E. R\'{\i}os}
 \affiliation{Departamento de Ingenier\'ia Qu\'imica, Universidad Tecnol\'ogica Nacional,
 	Facultad Regional Avellaneda, C.C. 1874, Argentina}

\begin{abstract}
	We analyze ground state (GS) factorization in general arrays of spins $s_i$ with $XXZ$
	couplings immersed in nonuniform fields. It is shown that an  exceptionally
	degenerate set of completely separable symmetry-breaking GS's can arise for a wide range of field configurations,
	at a quantum critical point where all  GS magnetization plateaus merge.
	Such configurations include alternating fields as well as zero bulk field solutions with edge
	fields only and intermediate solutions with zero field at specific sites, valid for
	$d$-dimensional arrays. The definite magnetization projected GS's at factorization can be analytically
	determined and depend only on the exchange anisotropies, exhibiting critical entanglement properties.
	We also show that some factorization compatible field configurations may result in field-induced
	frustration and nontrivial behavior at strong fields.
\end{abstract}

\maketitle

One of the most remarkable phenomena arising in finite interacting spin systems is that of {\it factorization}. For particular
values and orientations of the applied magnetic fields, the system possesses a {\it completely separable}
exact ground state (GS) despite the strong couplings existing between the spins. The close relation between
GS factorization and quantum phase transitions was first reported in \cite{KT.82} and has since been studied in
various spin models \cite{Mu.85,T.04,Am.06,KP.07,GI.07,GI.08,RCM.08,CRM.10,G.09,ARL.12,SC.13}, with general conditions
for factorization discussed in \cite{GI.08} and \cite{MRC.15}. Aside from some well known integrable cases
\cite{HB.31,RB.82,MT.99,S.99}, higher dimensional systems of arbitrary spin in general magnetic fields are not exactly solvable,
so that exact factorization points and curves provide a useful insight into their GS  structure.

The $XXZ$ model is an archetypal quantum spin system which has been widely studied
 to understand the properties of interacting many-body systems and their quantum phase transitions
 \cite{YY.66,JM.72,AM.95,DK.04,CR.06,TL.13}. It can emerge as an effective Hamiltonian in different scenarios,
 like bosonic and fermionic Hubbard models \cite{SP.12,GR.13,TC.14,Mlk.07}
and interacting atoms in a trapping potential
 \cite{SP.12, MV.16,Ha.17}. Renewed interest on it  has been enhanced by the recent advances in quantum control  
 with state-of-the-art technologies \cite{GA.14,Y.16},  which  enable  its finite size simulation even with tunable 
 couplings and fields in systems such as cold atoms in optical lattices \cite{SP.12,MV.16,Ha.17,DDL.03,Km.09,NB.13},  
 photon-coupled microcavities \cite{NA.17,CZZZG.10,BP.07}, superconducting Josephson junctions \cite{SS.15,W.17,BB.16,XL.14,DS.13}, 
 trapped ions  \cite{PC.04,BR.12,KK.12,AP.16,GA.14}, atoms on surfaces \cite{T.16} and quantum dots \cite{SO.11}. These features 
 make it a suitable candidate for implementing quantum information processing tasks 
 \cite{SP.12,MV.16,Ha.17,GA.14,Y.16,SO.11,LD.98,BK.98,BB.04,GN.09,BB.10,LB.11,st.09}. 

Our aim here is to show that in finite $XXZ$ systems of arbitrary spin under nonuniform fields,
highly degenerate exactly separable symmetry-breaking GS's can arise for a wide range of field
configurations in arrays of any dimension,
at an outstanding critical point where all magnetization plateaus merge and entanglement reaches full range.
The Pokrovsky-Talapov (PT)-type transition in a spin-$1/2$ chain in an alternating field \cite{AM.95} is
shown to correspond to this factorization.  Magnetization phase diagrams, showing non trivial behavior at
strong fields, and pair entanglement profiles  for distinct factorization compatible field configurations are presented,
together with analytic results for definite magnetization  GS's.

We consider an array of $N$  spins $s_i$ interacting through $XXZ$ couplings and immersed in a general
nonuniform magnetic field along the $z$ axis. The Hamiltonian reads
\begin{equation}
H=-\sum_i h^i S_i^z-\sum_{i<j}J^{ij}(S^x_i S^x_j+S^y_i S^y_j)+J_z^{ij}S^z_i S^z_j\,,\;\; \label{H}
\end{equation}
with $h^i$, $S_i^\mu$ the field and spin components at site $i$ and $J^{ij}$, $J_z^{ij}$ the exchange coupling strengths.
Since $H$ commutes with the total spin component $S^z=\sum_i S^z_i $, its eigenstates can be characterized by their total
magnetization $M$ along $z$. The exact GS will then exhibit definite $M$ plateaus as
the fields $h^i$ are varied, becoming maximally aligned ($|M|=S\equiv\sum_i s_i$) and hence completely separable
for sufficiently strong uniform fields. Otherwise it will be normally entangled.

We now investigate the possibility of $H$ having  a {\it nontrivial} completely separable GS of the form
\begin{equation}
|\Theta\rangle=\otimes_{i=1}^n e^{-\imath \phi_i S^z_i}e^{-\imath \theta_i S^y_i}|\uparrow_i\rangle=
|\nearrow\swarrow\nwarrow...\rangle\,,\label{Theta}
\end{equation}
where the local state $|\!\!\uparrow_i\rangle$ ($S^z_i|\!\!\uparrow_i\rangle=s_i|\!\!\uparrow_i\rangle$)
is rotated to an arbitrary direction $\bm{n}_{i}=(\sin\theta_i\cos\phi_i,\sin\theta_i\sin\phi_i,\cos\theta_i)$.
$|\Theta\rangle$  will be an exact eigenstate of $H$ iff two sets of conditions are met
\cite{MRC.15}. The first ones, 
\begin{eqnarray}
J^{ij}\cos\phi_{ij}(1-\cos\theta_i\cos \theta_j)&=&J^{ij}_z\sin\theta_i\sin\theta_j\,,\label{2}\\
J^{ij}\sin\phi_{ij}(\cos\theta_i-\cos \theta_j)&=&0\,,\label{1}
\end{eqnarray}
where $\phi_{ij}=\phi_i-\phi_j$, are field-independent and relate the alignment directions with the exchange couplings,
ensuring that $H$ does not connect $|\Theta\rangle$ with two-spin excitations. The second ones,
\begin{eqnarray}
h^i\sin\theta_i&=&{\textstyle\sum\limits_{j\neq i}}
s_j[J^{ij} \cos\phi_{ij}\cos\theta_i \sin\theta_j-J^{ij}_z\sin\theta_i \cos\theta_j],\label{3}\;\;\;\; \\
0&=&{\textstyle\sum\limits_{j\neq i}}s_j J^{ij}\sin\phi_{ij}\sin\theta_j\,,\label{4}
\end{eqnarray}
determine the {\it factorizing fields} (FF) and cancel all elements connecting $|\Theta\rangle$ with
single spin excitations, representing the mean field equations
$\partial_{\theta_i(\phi_i)}\langle\Theta|H|\Theta\rangle=0$.

\begin{figure}[t]
  \centering{\scalebox{.6}{\includegraphics{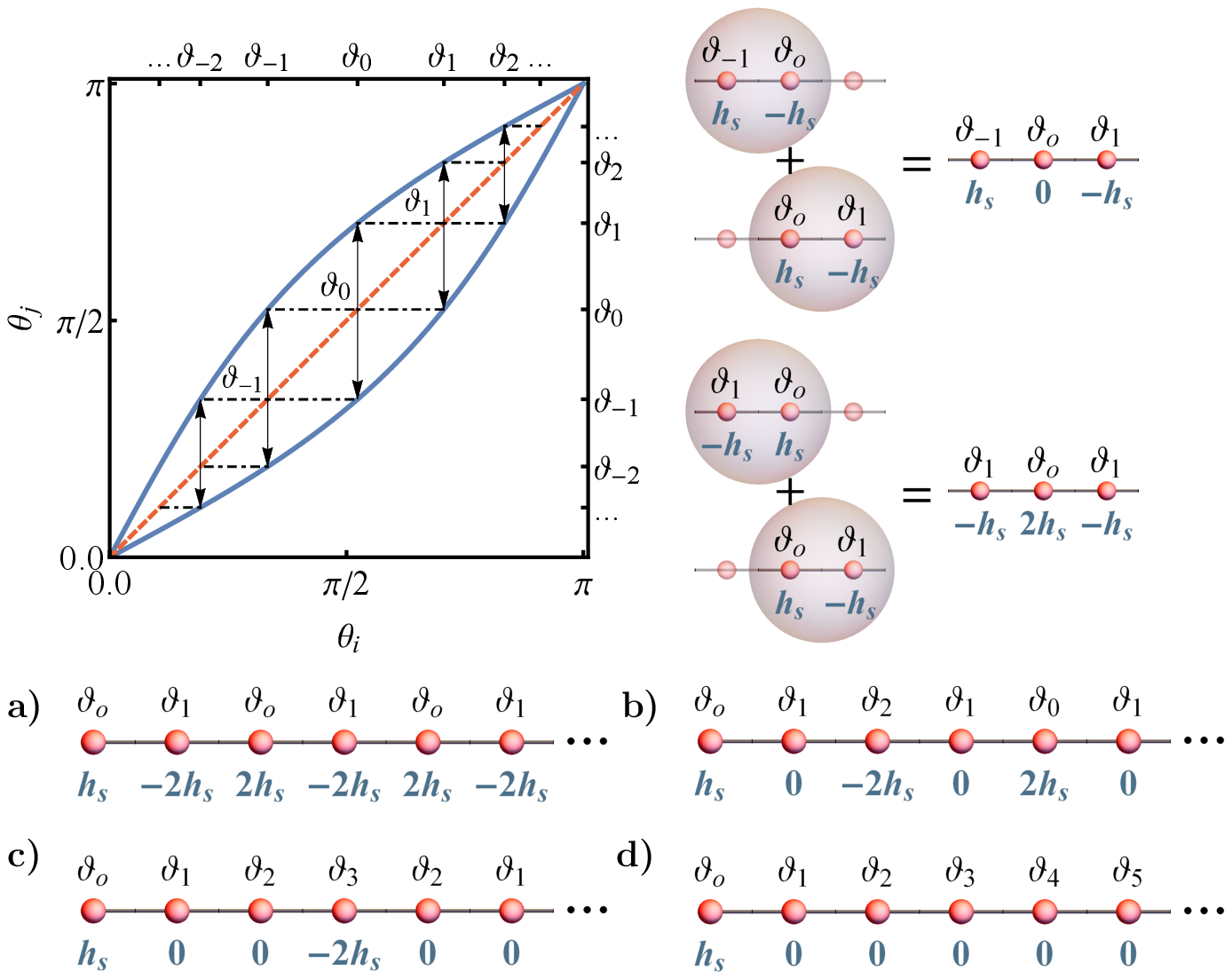}}} \vspace*{-0.25cm}
\caption{Top left: The two solutions of Eq.\ (\ref{t2}) for $\theta_j$ vs.\ $\theta_i$
	(thick solid lines).
For an arbitrary initial spin orientation $\vartheta_0$ at one site, successive application of Eq.\ (\ref{t2}) determines
the possible orientation angles (indicated by the arrows) of remaining spins in a factorized eigenstate
$|\Theta\rangle$. Each sequence of angles leads to a different factorizing field configuration determined by Eq.\ (\ref{hi}),
shown on the top right panels for 3 spins and on the bottom rows for the first 6 spins of a chain with uniform spin and couplings.
Two extremal cases arise: An alternating solution (a) and a zero bulk field
solution with edge fields only (d). Solutions with intermediate zero fields are also feasible (b,c).
In a cyclic chain the first field is $2h_{\rm s}$.}  \label{f1}
\end{figure}

These equations are always fulfilled by aligned states ($\theta_i=0$ or $\pi$ $\forall i$).  We now seek solutions
with $\theta_i\neq 0,\pi$ and $\phi_{ij}=0$ $\forall\, i,j$ \cite{note}.
Eqs.\ (\ref{1}) and (\ref{4}) are then trivially satisfied whereas Eq.\ (\ref{2}) implies
\begin{equation}
\eta_{ij}\equiv\frac{\tan(\theta_j/2)}{\tan(\theta_i/2)}=\Delta_{ij}\pm\sqrt{\Delta_{ij}^2-1}
\,,\label{t2}
\end{equation}
where $\Delta_{ij}=J_z^{ij}/J^{ij}=\Delta_{ji}$.
Such solutions  become then feasible if $|\Delta_{ij}|\geq 1$. For $|\Delta_{ij}|>1$
(\ref{t2}) yields two possible values of $\theta_j$ for a given $\theta_i$ ($\theta_j=\vartheta_{\pm 1}$
if $\theta_i=\vartheta_0$, see Fig.\ \ref{f1}, top left). And given $\theta_i,\theta_j$  $\neq 0,\pi$, there is a {\it single}
value $\Delta_{ij}=\frac{\eta_{ij}+\eta_{ij}^{-1}}{2}$
satisfying (\ref{t2}) ($\eta_{ij}^{-1}=\Delta_{ij}\mp\sqrt{\Delta^2_{ij}-1}$).

If Eq.\ (\ref{t2}) is satisfied for all coupled pairs, Eq.\ (\ref{3}) leads to the {\it factorizing fields}
\begin{equation}
h^i_{\rm s}=\sum_j  s_j\nu_{ij} J^{ij}\sqrt{\Delta^2_{ij}-1}\,,\label{hi}
\end{equation}
where $\nu_{ij}=-\nu_{ji}=\pm 1$ is the sign in (\ref{t2}). These fields are independent of the angles $\theta_i$
and always fulfill  the weighted {\it zero sum condition}
\begin{equation}\sum_i s_ih^i_{\rm s}=0\,.\label{zs}\end{equation}
The ensuing energy $E_\Theta=-\sum_i s_i\bm{n}_i\cdot[\bm{h}^i_{\rm s}+\sum_{j>i} {\cal J}^{ij}s_j\bm{n}_{j}]$
(${\cal J}^{ij}_{\mu\nu}\equiv J^{ij}_{\mu}\delta_{\mu\nu}$) depends {\it only on the strengths}  $J_z^{ij}$:
\begin{equation}
E_\Theta =-\sum_{i<j}s_i s_j J_z^{ij}\,,\label{Et}
\end{equation}
coinciding with that of the $M=\pm S$  aligned states in such field.
It is proved (see Appendix A) that if $J_z^{ij}\geq 0$  $\forall i,j$,
 (\ref{Et}) {\it is the GS energy} of such $H$. Essentially, $H$ can
 be written as a sum of pair Hamiltonians $H^{ij}$ whose GS energies are precisely $-s_i s_j J_z^{ij}$.
 If $J_z^{ij}<0$ $\forall i,j$, it is instead its  highest eigenvalue.

These separable eigenstates do not have a definite magnetization, breaking the basic symmetry of $H$
and containing components with all values of $M$. They can then only arise
at an exceptional point where the GS becomes $2S+1$ degenerate and  {\it all GS magnetizations plateaus  coalesce}:
Since $[H,P_M]=0$, with $P_M=\frac{1}{2\pi}\int_0^{2\pi} e^{i\varphi(S^z-M)}d\varphi$  the projector onto total
magnetization $M$,  $HP_M|\Theta\rangle=E_{\Theta}P_M|\Theta\rangle$ for all  $M=-S,\ldots,S$. All components of
$|\Theta\rangle$ with definite $M$ are exact
eigenstates with the same energy (\ref{Et}). Moreover,  the normalized projected states are {\it independent} of
both $\phi$ and the seed angle $\theta_1=\vartheta_0$,
depending just on the exchange anisotropies $\Delta_{ij}$ and
the signs $\nu_{ij}$ (see Appendix A):
\begin{eqnarray}\!\!\!\!\!P_M|\Theta\rangle
&\propto&\!\!\!\sum_{m_1,\ldots,m_N\atop
	\sum_i m_i=M}[\prod_{i=1}^N\sqrt{{\textstyle\binom{2s_i}{s_i-m_i}}}
\eta_{i,i+1}^{\sum_{j=1}^{i}m_j}]|m_1\ldots m_N\rangle
\label{PM2},\;\;\;\;
\end{eqnarray}
where $\eta_{i,i+1}$ denote  the ratios (\ref{t2}) along {\it any} curve in the array joining all
coupled spins. In contrast with $|\Theta\rangle$, these states are entangled $\forall\,|M|\leq S-1$
and represent the actual limit of the exact GS along the $M^{\rm th}$ magnetization plateau  as the
factorization point is approached.

As a basic example, for a {\it single spin-$s$ pair} with $J^{ij}=J$, GS factorization will arise whenever $J_z>0$
and $|\Delta|=|\frac{J_z}{J}|> 1$,  at opposite FF
 $h^1_{\rm s}=-h^2_{\rm s}=\pm h_{\rm s}$, with
 \begin{equation}h_{\rm s}=sJ\sqrt{\Delta^2-1}\,.\label{hs}\end{equation}
  At these points the GS is $4s+1$ degenerate, with energy $E_{\Theta}=-s^2 J_z$ and projected GS's
\begin{eqnarray}{\textstyle\frac{P_M|\Theta\rangle}{\sqrt{\langle\Theta|P_M|\Theta\rangle}}}
&=&\sum_m{\textstyle\sqrt{\frac{\binom{2s}{s-m}\binom{2s}{s+m-M}}{Q_{2s-M}^{M,0}(\eta)}}
	\eta^{s+m-M}|m,M-m\rangle\,},\label{epm2}\;\;\;\;\;\;\;
\end{eqnarray}
where $Q_{n}^{m,k}(\eta)=(\eta^2-1)^nP_{n}^{m-k,m+k}(\frac{\eta^2+1}{\eta^2-1})$ with
$P_n^{\alpha,\beta}(x)$ the Jacobi Polynomials and $\eta$ the ratio (\ref{t2}). These states are
entangled, with (\ref{epm2}) their Schmidt decomposition.

{\it Spin chains}. The factorized GS's of a single pair can be used as building blocks for constructing
separable GS's of a chain of $N$ spins (Fig.\ \ref{f1}). For
first neighbor couplings, after starting with a seed
$\theta_1=\vartheta_0\in(0,\pi)$ at the first spin, $\theta_2,\ldots,\theta_N$ are determined by Eq.\ (\ref{t2}).
The two choices for $\theta_j$ at each step then lead  to
$2^{N-1}$ distinct factorized states and FF configurations in an open chain.

For {\it uniform} spins $s_i=s$ and couplings $J^{i,i+1}=J$, $\Delta_{i,i+1}=\Delta$ $\forall i$,
the FF (\ref{hi})  become $h^i_{\rm s}=\nu_i h_{\rm s}$,
with $h_{\rm s}$ given by (\ref{hs}) and $\nu_i={\sum_j} \nu_{ij}=\pm 2$ or $0$ for bulk spins and $\pm 1$ for
edge spins. Among the plethora of factorizing spin and field configurations, two extremal cases stand out:
 A N\'eel-type configuration
$\vartheta_0\vartheta_1\vartheta_0\vartheta_1\ldots$, implying an alternating field $h_{\rm s}^i=\pm 2(-1)^{i}h_{\rm s}$
for bulk spins and $|h_{\rm s}^1|=|h_{\rm s}^N|=h_{\rm s}$ for edge spins  (Fig.\ \ref{f1} a), and a solution with increasing angles
$\vartheta_0,\vartheta_1,\vartheta_2,\ldots$, implying
zero bulk field and edge fields $h_{\rm s}^1=-h_{\rm s}^N=\pm h_{\rm s}$ (d). Solutions with intermediate
 zero fields are also feasible (b,c). In a {\it cyclic} chain ($N+1\equiv 1$,
 $J_\mu^{1N}=J_\mu$) the number of configurations is smaller, i.e.\ $\binom{N}{N/2}$
($\approx \frac{2^{N-1}}{\sqrt{\pi N/8}}$ for large $N$),
as (\ref{t2}) should be also fulfilled for the 1--$N$ pair, entailing $\theta_{N}=\vartheta_{\pm 1}$,  $N$ {\it even}
and an equal number of positive and negative choices in (\ref{t2}). For
$\Delta\rightarrow 1$, $h_{\rm s}\rightarrow 0$ and all solutions converge to a uniform $|\Theta\rangle$ 
($\theta_i$ constant, Eq.\ (\ref{t2})).

\begin{figure}[t]
\centering{{\scalebox{.75}{\includegraphics{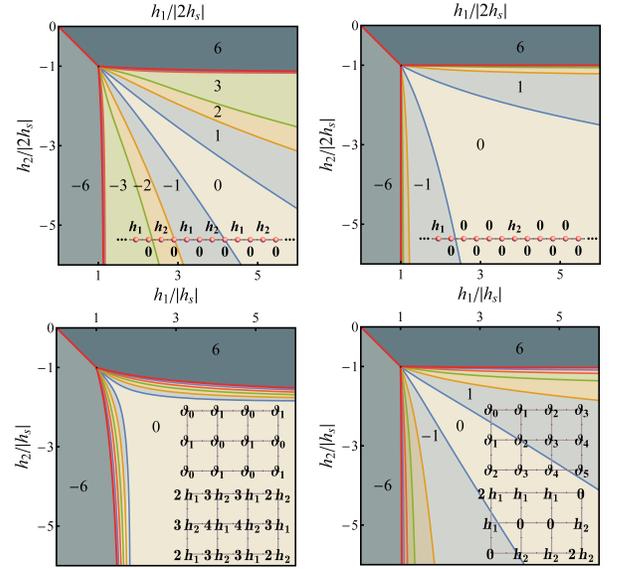}}}}
\vspace*{-0.cm}
	\caption{Exact GS magnetization diagram for distinct spin arrays and field configurations
		with $\Delta=1.2$. Top: Cyclic $N=12$
		spin-$1/2$ chain with next alternating fields (left) and zero bulk field (right). Bottom: Open $3\times 4$ spin-$1/2$
		arrays with alternating (left) and zero bulk (right) field configurations. All plateaus merge at the factorizing point,
		where the GS has the indicated angles. Field induced frustration in configurations
 with zero fields lead to a reduced $M=0$ plateau.}

	\label{f4}
\end{figure}

{\it Spin lattices}. Previous arguments can be extended to $d$-dimensional spin arrays, like spin-star geometries 
\cite{st.09} and square or cubic lattices with first neighbor couplings and fixed $\Delta_{ij}=\Delta$. 
As the angles $\theta_j$ of all spins coupled to spin $i$ should satisfy (\ref{t2}), they must differ from $\theta_i$ 
in just one step: $\theta_j=\vartheta_{k\pm 1}$ if $\theta_i=\vartheta_k$ (Fig.\ \ref{f1}). Nonetheless, the number 
of feasible spin and field configurations still increases exponentially with lattice size (see Appendix E for a detailed discussion). 
The FF are $h^{\bm i}_{\rm s}=\pm\nu_{\bm{i}}h_s$ with $\nu_{\bm i}$ integer.
In particular, the previous two extremal solutions remain feasible
(see Fig.\ \ref{f4}): By choosing in (\ref{t2}) {\it alternating} signs along rows, columns, etc.\
we obtain  {\it alternating} FF $h^{\bm{i}}_{\rm s}=\pm 2d h_{\rm s}$ for bulk spins
($h^{ij}_{\rm s}=\pm 4(-1)^{i+j}h_{\rm s}$ for $d=2$), with smaller values at the borders. And by always choosing
 the {\it same} sign in (\ref{t2}), such that $\vartheta$ increases along rows, columns, etc.\ the FF will be
 {\it zero} at {\it all} bulk spins, with nonzero fields $\nu_i h_{\rm s}$ just at the border.

{\it Definite $M$ reduced states}. For uniform anisotropy $\Delta$,
all ratios $\eta_{i,i+1}$ in the projected states (\ref{PM2}) will be either $\eta$ or $\eta^{-1}$
and more explicit expressions can be obtained. For instance, for a spin-$s$ array in an alternating FF,
Eq.\ (\ref{PM2}) leads, in any dimension, to just three distinct reduced pair states
$\rho^M_{ij}$ of spins $i\neq j$: $\rho^M_{oe}$ (odd-even), $\rho^M_{oo}$ and $\rho^M_{ee}$,
which {\it will not depend on the actual separation between the spins} since $\rho^M_{i,j+k}=\rho^M_{i,j}$
$\forall$ $k$ even, due to the form of $|\Theta\rangle$.
 Their nonzero elements are
 \begin{equation}
(\rho^M_{ij})^m_{m_j,m'_j}=
{\eta^{f_{ij}}\frac{\sqrt{C^{s,m}_{m_j}C^{s,m}_{m'_j}}Q_{Ns-2s-M+m}^{M-m,(\delta+2l_{ij})s}(\eta)}
{Q_{Ns-M}^{M,\delta s}(\eta)}}
\label{redp},
\end{equation}
where $m=m_i+m_j=m'_i+m'_j$ is the pair magnetization
($[\rho^M_{i,j},S^z_i+S^z_j]=0$), $Q_{n}^{m,k}(\eta)$ was defined in (\ref{epm2}), $C^{s,m}_k=\binom{2s}{s-k}\binom{2s}{s-m+k}$
and $f_{ij}=2s-m_j-m'_j,0,4s-2m$, $l_{ij}=0,-1,1$ for
$oe,oo,ee$ pairs, with $\delta=0(1)$ for $N$ even (odd). For $|M|<Ns$, these states are mixed (implying entanglement
with the rest of the array) and also entangled for finite $N$, entailing that pair entanglement will reach {\it full range},
as discussed below.
\begin{figure}[t]
	\centering{{\scalebox{.5}{\includegraphics{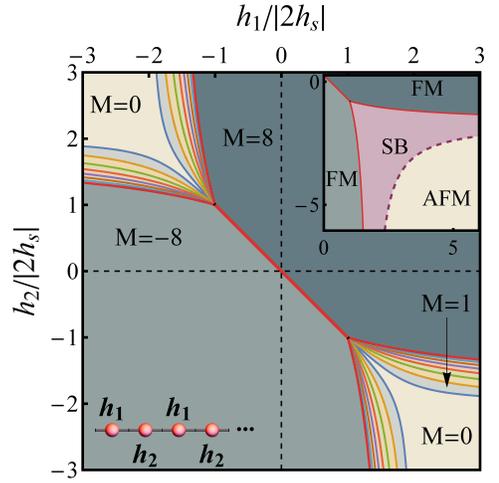}}}}
\caption{GS magnetization diagram for alternating fields $h^{2i-1}=h_1$, $h^{2i}=h_2$ in an $N=8$ spin 1
	$XXZ$ chain with $\Delta=1.2$. All magnetization plateaus $M=Ns,\ldots,-Ns$ coalesce at the factorizing fields $
	h_1=-h_2=\pm 2h_{\rm s}$.  The inset indicates the mean field (MF) phases.}
	\label{f2}
\end{figure}

{\it Magnetic Behavior}. The  FF (\ref{hi}) are {\it critical points} in the multidimensional field space
$(h^1,\ldots,h^N)$, as seen in Fig.\ \ref{f2} for a finite spin $1$ cyclic chain in an alternating field $(h_1,h_2,h_1,\ldots)$.
While a large part of the field plane $(h_1,h_2)$ corresponds for $\Delta>1$  to an aligned  GS ($M=\pm Ns$),  sectors with GS
magnetizations $|M|<Ns$ emerge precisely at the FF $h_1=-h_2=\pm 2h_{\rm s}$.  These fields coincide
with those of the PT-type transition for $h_1=-h_2$ in a spin $1/2$ chain \cite{AM.95}, which then corresponds to present
GS factorization (holding for {\it any} spin $s$). The border of the aligned sector is actually determined by the hyperbola branches
\begin{equation}{\textstyle(\frac{h_1}{2sJ}\pm \Delta)(\frac{h_2}{2sJ}\pm \Delta)=1} \,,\;\;
\label{hyp}
\end{equation}
(for $|h_i|>2h_s$, $\mp \frac{h_i}{2sJ}<\Delta$,  see Appendix C), which {\it cross} at the FF if $\Delta\geq 1$.
Eq.\ (\ref{hyp}) also determines the onset of the symmetry-breaking (SB) MF solution (inset in Fig.\ \ref{f2}), which ends in an
antiferromagnetic (AFM) phase for strong fields of opposite sign (Appendix C). 

Along lines $h_2=h_1+\delta$, the exact GS for $\Delta>1$ then undergoes a single $-Ns\rightarrow Ns$ transition
if $\delta<|4h_{\rm s}|$ but $2Ns$ transitions $M\rightarrow M+1$ if $|\delta|>4h_{\rm s}$, starting at the border (\ref{hyp}).
Hence, at factorization, application of further fields $(\delta h_1,\delta h_2)=\delta h(\cos\gamma,\sin\gamma)$ enables
to select {\it any} magnetization plateau, which initially emerge at straight lines at angles
$\tan\gamma_{M}=\frac{\langle S_1^z\rangle_{M}-\langle S_1^z\rangle_{M-1}}{\langle S_2^z\rangle_{M-1}-\langle S_2^z\rangle_{M}}$ \cite{Szm}. 
Moreover, at this point an additional {\it arbitrarily oriented} local field $\bm{h}^i$ applied at site  $i$
will {\it bring down a single separable GS} (that with $\bm{n}_i\parallel \bm{h}^i$), splitting the $2Ns+1$
degeneracy and enabling  a {\it separable GS engineering} (Appendix B). 

\begin{figure}[t]
	\centering{{\scalebox{.75}{\includegraphics{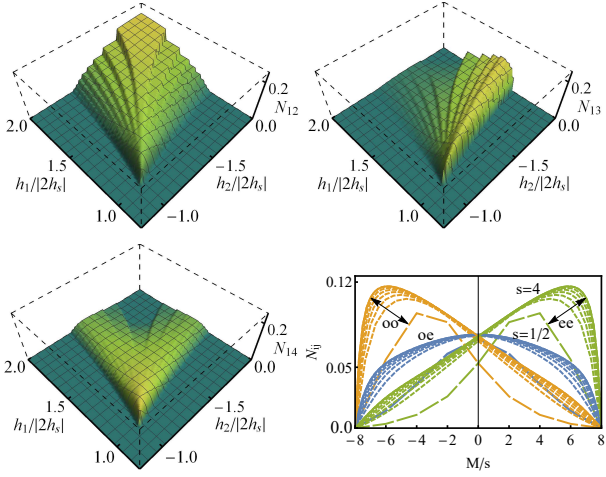}}}}
	\vspace*{-0.cm}
	\caption{Exact pair negativities $N_{ij}$ between spins $i$ and $j$
	in the exact GS of the spin-$1$ chain of Fig.\ \ref{f2},
		for fields $h_1,h_2$ of opposite sign and first (top left), second (top right) and
		third (bottom left) neighbors.
		Bottom right:
		The exact pair negativities at factorization ($h_1=-h_2=2h_{\rm s}$) in the definite magnetization GS's,
		for identical $N=8$ spin-$s$ chains with $s=1/2,\ldots,4$. At this point
 there are just three distinct pair negativities: $N_{oe}$ (odd-even), $N_{oo}$ and $N_{ee}$,
independent of the actual separation $|i-j|$ and dependent on $M$.}
			\label{f3}
\end{figure}

The entanglement between two spins $i,j$ in the same chain is depicted  in Fig.\ \ref{f3} through the
pair negativity $N_{ij}=({\rm Tr}\,|\rho^{\rm pt}_{ij}|-1)/2$ \cite{Neg}, where $\rho_{ij}^{\rm pt}$ is
the partial transpose of $\rho_{ij}$. $N_{ij}$ exhibits a stepwise behavior, reflecting the magnetization
plateaus, with the onset of entanglement determined precisely by the FF and that of the $|M|=Ns-1$ plateau
(Eq.\ (\ref{hyp})). Due to the interplay between fields and exchange couplings, $N_{ij}$ increases for
decreasing $|M|$ for contiguous pairs (top left), since the spins become less aligned, but shows an asymmetric behavior
for second neighbors (top right), as these pairs become more aligned when $M$ increases and acquires the same sign as the
corresponding field. Third neighbors (bottom left) remain appreciably entangled
at the FF, since there $N_{14}=N_{12}=N_{oe}$. This property also holds at the border (\ref{hyp}) due to 
the $W$-like structure of the $M=Ns-1$ GS (see Appendix C for  expressions of $N_{ij}$ and the concurrence). 
The exact negativities at factorization in the projected
states (\ref{PM2}) (bottom right), obtained from (\ref{redp}),
exhibit the same previous behavior with $M$ for any $s$. They are in compliance with the monogamy property,
decreasing as $N^{-1}$ for large $N$ at fixed finite $M$.

The general picture for other field configurations is similar, but
differences do arise, as shown in Fig.\ \ref{f4}.  While in all cases the $|M|<Ns$  plateaus emerge from the FF,
 with the diagram  of the alternating square lattice remaining similar to that of Fig.\ \ref{f2},
 the chain with next alternating fields ($h_1,0,h_2,0,\ldots$) exhibits a much reduced $M=0$ plateau and 
wider sectors with finite $|M|\leq Ns/2$. This effect is due to the intermediate spins with zero
field, which are frustrated for $M=0$  ({\it field induced frustration})
and become more rapidly aligned with the stronger field as it increases, and  facilitates the selection
through nonuniform fields of different magnetizations. A similar though attenuated effect occurs in the zero-bulk
field configurations (right panels). Moreover, in these three cases selected pairs of spins with zero field can remain 
significantly entangled in the $M=0$ plateau for strong  $h_1$, $h_2$ of opposite signs,
as shown in Appendix D.
The definite $M$ states at factorization become more complex, leading to several distinct reduced pair states,
whose negativities become maximum at different $M$ values (Appendix D).

We have proved the existence of a whole family of completely separable symmetry breaking exact GS's
in arrays of general spins with $XXZ$ couplings, which arise for a wide range of nonuniform field configurations of
zero sum (Eq.\ (\ref{zs})). They correspond to a multi-critical
point where {\it all}  GS magnetization plateaus coalesce, and
where entanglement reaches full range for all nonaligned definite-$M$ GS's. This point can arise even for simple field architectures
like just two nonzero edge fields of opposite sign in a  chain or edge fields in a lattice, and for {\it any} size $N\geq 2$
and spin $s\geq 1/2$. Different GS magnetization diagrams can be generated, opening  the possibility to access distinct types of  GS's
(from separable with arbitrary spin orientation at one site to entangled with any $|M|<S$) with small field variations,
 and hence to engineer specific GS's useful for quantum processing tasks.  Recent tunable realizations of finite $XXZ$ arrays 
\cite{MV.16,Ha.17,XL.14} (see also Appendix F) 
 provide a promising scenario for applying these results.  
\begin{acknowledgments}
	The authors acknowledge support from CONICET (MC, NC) and CIC (RR) of Argentina.
	Work supported by CONICET PIP 11220150100732.
\end{acknowledgments}
\begin{center}
\section*{\it Supplemental Material}
\end{center}
\section{Appendix A. Proof of ground state condition}

We show here that the factorized state $|\Theta\rangle=|\theta_1\theta_2\ldots\rangle$,
with angles $\theta_i$ satisfying Eq.\ (7)  and $\phi_{ij}=0$,  is a ground state (GS) of the
Hamiltonian (1) if $J_z^{ij}>0$ for all coupled pairs  and if the fields are given by Eq.\  (8).  \\
{\it Proof:} We first consider a single pair $i\neq j$ ($N=2$).
We set  $\phi_i=\phi_j=0$ as its value will
not affect the average energy ($[H,S^z]=0$). The factorized pair state 
$|\theta_i\theta_j\rangle$ has
in the standard basis the explicit form
\begin{equation}\tag{A1}
|\theta_i\theta_j\rangle=\!\!\!\mathop{\otimes}_{k=i,j}
\!\sum_{m_k=-s_k}^{s_k}\!\!\!\!\!\!{\textstyle\sqrt{(^{\;\;\;2s_k}_{s_k-m_k})}
	\sin^{s_k-m_k}\!\frac{\theta_k}{2}\cos^{s_k+m_k}
	\!\frac{\theta_k}{2}|m_k\rangle}\,,\label{MFs}
\end{equation}
where $S_k^z|m_k\rangle=m_k|m_k\rangle$. For $J^{ij}_z>0$ and $J^{ij}>0$,
Eq.\ (7)  admits solutions with $\theta_i,\theta_j\in(0,\pi)$,
in which case all coefficients in the expansion (\ref{MFs}) are strictly positive.
Therefore,  $|\theta_i\theta_j\rangle$ must be a GS of the pair Hamiltonian $H^{ij}$
if the fields satisfy (8), since it is an exact eigenstate and since for $J^{ij}>0$, all nonzero off-diagonal elements
of $H^{ij}$ in this basis are negative (implying that  $\langle H^{ij}\rangle$ can always be
minimized by a state with all coefficients of the same sign in this basis, which
cannot be orthogonal to $|\theta_i\theta_j\rangle$).

A rotation of angle $\pi$ around the $z$ axis of one of the spins (say $j$) will change the sign of $J^{ij}$
and $\theta_j$, leaving $J_z^{ij}$, the fields and the spectrum of $H^{ij}$ unchanged.
Thus, $|\theta_i,-\theta_j\rangle$ will be a GS of such $H^{ij}$,
with $\theta_i,-\theta_j$  satisfying (7) for $\Delta_{ij}<0$ (with the sign $\nu_{ij}\rightarrow-\nu_{ij}$) and the same 
fields still satisfying (8).

Previous arguments also show that for any sign of $J^{ij}$,  $|\theta_i\theta_j\rangle$ will be the state with
the {\it highest} eigenvalue of $H^{ij}$  if  $J_z^{ij}<0$,   since  it will be the GS of $-H^{ij}$.

Considering now a general array of spins with Eq.\ (7) satisfied by all coupled pairs and the fields given by (8),
we may write the full $H$ as $\sum_{i<j}H^{ij}$, with
\begin{equation}\tag{A2}
H^{ij}=-h_{\rm s}^{ij} S_i^z-h^{ji}_{\rm s}S_j^z-J^{ij}(S_i^xS_j^x+S_i^yS_j^y)-J^{ij}_zS_i^zS_j^z\,,
\end{equation}
and $h^{ij}_{\rm s}=
\nu_{ij}s_jJ^{ij}\sqrt{\Delta_{ij}^2-1}=-s_jh^{ji}_{\rm s}/s_i$ the factorizing fields
for the pair $i,j$ ($\sum_j h^{ij}_{\rm s}=
h^i_{\rm s}$).  Then, for $J_z^{ij}>0$ $\forall$ $i,j$, $|\Theta\rangle$ will be the GS of $H$ since
$|\theta_i\theta_j\rangle$ will be the GS of each $H^{ij}$. By the same arguments,
if $J_z^{ij}<0$ $\forall$ $i,j$ such state will be that with the highest eigenvalue of $H$.
\qed

From the form of Eq.\ (\ref{MFs}),
it is seen that the projected states $P_M|\Theta\rangle$ will acquire the form
of Eq.\ (11)   when Eq.\ (7) is satisfied for all coupled pairs.

\section{Appendix B. Separable ground state extraction}

\begin{figure}[t]
	\centering{\scalebox{.8}{\includegraphics{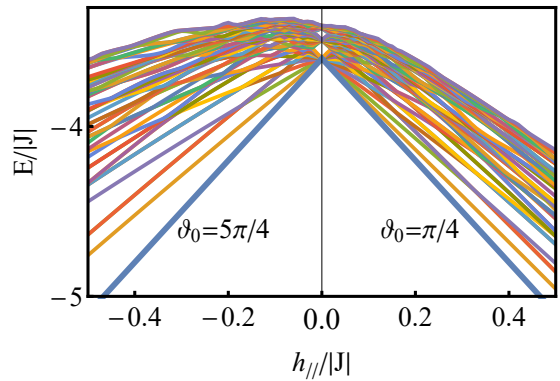}}}
	\vspace*{-0.0cm}
	\caption{Energy spectrum of the lowest $50$ eigenstates of a spin-$1/2$ $XXZ$ cyclic chain of $N=12$
		spins with uniform $\Delta=1.2$ in alternating factorizing fields when additional local fields (\ref{c1})
		with $\bm{n}=(\sin(\pi/4),0,\cos(\pi/4))$  are applied at odd sites.
		The lowest thick line represents the energy of the extracted separable GS. At $h_\parallel=0$,
		$|\Theta\rangle$ is $2Ns+1$-fold degenerate, while for $h^i_\parallel >0$ $(<0)$ the GS is
		nondegenerate with spins at odd sites pointing along $\theta_i=\pi/4$ $(-3\pi/4)$.}
	\label{fsm1}
\end{figure}
\vspace*{-0.25cm}

Given a separable eigenstate $|\Theta\rangle$, the addition at a given site $i$ of a local magnetic field
$\bm{h}^i_\parallel=h^i_\parallel \bm{n}_i$ parallel to the spin alignment direction $\bm{n}_i$ will just shift
its energy by $-s h^i_\parallel$ \cite{MRC.15}. In the present case, at factorization the angles $\phi_i$ and $\theta_i$ at a given site
can be arbitrarily chosen, since it can be considered a seed site.  Thus,  a local field
\begin{equation}\tag{B1}
\bm{h}^i=\bm{h}^i_{\rm s}+\bm{h}^i_\parallel\,,
\label{c1}
\end{equation}
with $\bm{h}^i_{\rm s}$ the transverse factorizing field and $\bm{h}^i_\parallel$ {\it  any arbitrarily
	oriented} local field, will select a separable  GS $|\Theta\rangle$ with $\bm{n}_i\parallel\bm{h}^i$,
lowering its energy and thereby splitting the $2Ns+1$ degeneracy, bringing down a nondegenerate separable GS (Fig.\ \ref{fsm1}).
Note that no other eigenstate will decrease its energy faster with $|\bm{h}^i_{\parallel}|$ than this $|\Theta\rangle$. The GS  energy
can be further lowered by means of additional local fields $\bm{h}^j_{\parallel}$ at sites $j$ 
along directions $\bm{n}_j$ compatible with this $|\Theta\rangle$.
As illustration of this effect, Fig.\ \ref{fsm1} depicts
the energy spectrum of a cyclic spin chain of $N=12$ spins with uniform first-neighbor 
couplings in an alternating factorizing field configuration when local fields (\ref{c1}) 
are applied at odd sites (with the $\bm{h}^i_{\rm s}$ fixed). 
Such separable states could be useful for initialization and for quantum annealing protocols.

\section{Appendix C. The $\bm{M=\pm (Ns-1)}$ GS for an alternating field and the onset of entanglement}
For any dimension $d$, the exact GS with magnetization $M=Ns-1$ of a spin-$s$ 
array in an alternating field with cyclic uniform $XXZ$ couplings is necessarily of the form
\begin{equation}\tag{C1}
|Ns-1\rangle=\cos\alpha |W_{o}\rangle+\sin\alpha|W_{e}\rangle\,,\label{GSs}
\end{equation}
where $|W_{^o_e}\rangle=\frac{1}{\sqrt{Ns}}\sum_{i\,^{\rm odd}_{\rm even}} S_{i}^-|Ns\rangle$
are $W$-like states for odd
and even spins  ($S_i^-=S_i^x-iS_i^y$) and $|Ns\rangle$ denotes the aligned $M=Ns$ state.
The angle $\alpha$ can be obtained from
the diagonalization of $H$ in the two-dimensional subspace spanned by the states
$|W_{^o_e}\rangle$, where
$\langle W_{^o_e}|H|W_{^o_e}\rangle=E_{Ns}+2sJ_z+ h_{^1_2}$ and
$\langle W_o|H|W_e\rangle=-2sJ$, with
\begin{equation}\tag{C2}
E_{Ns}={\textstyle-Ns(\frac{h_1+h_2}{2}+sJ_z)}\,,\end{equation}
the energy of the aligned state.  We then obtain
\begin{equation}\tag{C3}
E_{Ns-1}= E_{Ns}+{\textstyle\frac{h_1+h_2}{2}+2sJ_z-\sqrt{(\frac{h_1-h_2}{2})^2+(2sJ)^2}},\label{C1}\end{equation}
for the lowest $M=Ns-1$ energy, with
$^{\cos\alpha}_{\sin\alpha}=\sqrt{\frac{\lambda\mp (h_1-h_2)/2}{2\lambda}}$
and $\lambda=\sqrt{(\frac{h_1-h_2}{2})^2+(2sJ)^2}$. Similar expressions with $h_i\rightarrow-h_i$ hold for $E_{-Ns}$ 
and $E_{-Ns+1}$.

The $Ns\rightarrow Ns-1$ GS transition occurs for fields $(h_1,h_2)$ satisfying $E_{Ns}=E_{Ns-1}$, which leads to the
upper expression in Eq.\ (15). The
$-Ns\rightarrow -Ns+1$ GS transition is similarly obtained replacing $h_i$ by $-h_i$ and leads to the lower expression in (15).
These transitions determine the onset of GS entanglement.

While the previous exact GS transitions are sharp, at the border any linear combination of $|Ns\rangle$ and $|Ns-1\rangle$
is also a GS, including  $|Ns\rangle+\varepsilon|Ns-1\rangle$, which, up to first order in $\varepsilon$,  is a symmetry-breaking (SB)
product state $|\Theta\rangle$ with $\sin\frac{\theta_{i}}{2}\propto \varepsilon\cos\alpha\,(\sin\alpha)$ for odd (even) $i$.
Therefore, the onset of the SB mean field phase
coincides here with the exact onset of the $|M|=Ns-1$ GS, given by the hyperbola branches (15). The SB MF state
(obtained from Eqs.\ (5)--(6) at fixed $h^i$) is in fact a N\'eel-type state with
\begin{equation}\tag{C4}\cos\theta_{o(e)}={\textstyle\frac{sJ}{2h^2_{\rm s}}\left[h_{e(o)}\Delta-h_{o(e)}
	\sqrt{\frac{h_{e(o)}^2-4h_{\rm s}^2}{h_{o(e)}^2-4h_{\rm s}^2}}\,\right]}\,.\end{equation}
This phase extends into the $M=0$ plateau and ends in an antiferromagnetic (AFM) phase $\langle S^z_i\rangle=\pm s(-1)^{i}$ (inset of Fig.\ 2),
which is the lowest MF phase  for fields satisfying
\begin{equation}\tag{C5} (\frac{h_1}{2sJ}\pm \Delta)(\frac{h_2}{2sJ}\mp\Delta)\leq -1\;\;{\rm if}\;\;\mp\frac{h_1}{2sJ}> \Delta\,.\end{equation}

The reduced mixed state of a spin pair $i,j$ in the exact $M=Ns-1$ GS (\ref{GSs}) will  be independent
of separation for similar pairs,
i.e., odd-odd (oo), odd-even (oe) and even-even (ee) pairs, and will commute with the pair
magnetization $S^z_i+S^z_j$. Its only nonzero matrix elements will be those for $m=2s$ and $2s-1$.
The nonzero block in the subspace spanned by $\{|\uparrow\uparrow\rangle,
|\uparrow\downarrow\rangle,|\downarrow\uparrow\rangle\}$ will be
\[\rho_{oe}=\begin{pmatrix}1-\frac{2}{N}&0&0\\0&\frac{2}{N}\sin^2\alpha&\frac{1}{N}\sin 2\alpha\\0&
\frac{1}{N}\sin2\alpha&\frac{2}{N}\cos^2\alpha
\end{pmatrix},
\]
for odd-even pairs and
\[\rho_{oo}=\begin{pmatrix}1-\frac{4}{N}\cos^2\alpha&0&0\\0&
\frac{2}{N}\cos^2\alpha&\frac{2}{N}\cos^2\alpha \\0&\frac{2}{N}\cos^2\alpha&\frac{2}{N}\cos^2\alpha
\end{pmatrix},
\]
for odd-odd pairs, with $\cos^2\alpha\rightarrow \sin^2\alpha$ for even-even pairs ($\rho_{ee}$).
They are spin-independent for fixed $\alpha$, i.e., fixed scaled couplings $sJ_\mu$. The exact negativities read
\begin{align}
N_{oe}&={\textstyle\sqrt{(\frac{1}{2}-\frac{1}{N})^2+\frac{\sin^2 2\alpha}{N^2}}-(\frac{1}{2}-\frac{1}{N})}, \tag{C6}\\
N_{oo}&={\textstyle\sqrt{(\frac{1}{2}-\frac{2\cos^2\alpha}{N})^2+\frac{4\cos^2\alpha}{N^2}}-(\frac{1}{2}-
	\frac{2\cos^2\alpha}{N})}, \tag{C7}
\end{align}
with $\cos^2\alpha\rightarrow\sin^2\alpha$ for $N_{ee}$. For large $N$,
$N_{ij}\approx \frac{1}{4}C^2_{ij}$, where
\begin{align}
C_{oe}&=\frac{2|\sin 2\alpha|}{N}=\frac{2}{N}{\textstyle \frac{2s|J|}{\sqrt{(\frac{h_1-h_2}{2})^2+(2sJ)^2}}}
\,,\tag{C8}\\
C_{oo}&=\frac{4\cos^2\alpha}{N}=\frac{2}{N}(1-{\textstyle
	\frac{(h_1-h_2)/2}{\sqrt{(\frac{h_1-h_2}{2})^2+(2sJ)^2}}})\,,\tag{C9}\\
C_{ee}&=\frac{4\sin^2\alpha}{N}=\frac{2}{N}(1+{\textstyle
	\frac{(h_1-h_2)/2}{\sqrt{(\frac{h_1-h_2}{2})^2+(2sJ)^2}}}),\tag{C10}
\end{align}
are the associated concurrences. At factorization these results coincide with those derived from (14) for $M=Ns-1$.
The similar expressions for $M=-Ns+1$ are obtained replacing $h_i$ by $-h_i$. It is then seen that for large positive odd
fields $h_1$ and $M=Ns-1$, both $C_{oo}$ and $C_{oe}$ vanish
while $C_{ee}\rightarrow 4/N$ ($W$-state result for $N/2$ spins), whereas for large negative even fields $h_2$ and $M=-Ns+1$,
$C_{oo}\rightarrow 4/N$ while both $C_{oe}$ and $C_{ee}$ vanish, in agreement with the behavior seen in Fig.\ 3.

\section{Appendix D. Pair negativities}
\begin{figure}[t]
	\centering{\scalebox{.61}{\includegraphics{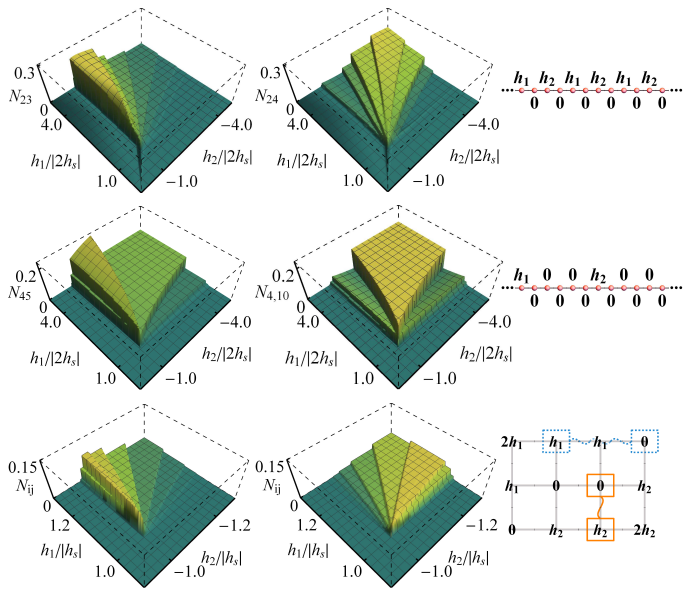}}}
	\vspace*{-0.0cm}
	\caption{Pair negativities $N_{ij}$ in the exact GS of an $N=12$ spin-$1/2$ cyclic chain with next alternating
		fields (top row) and zero bulk fields (center row), and an open $3\times 4$ spin-$1/2$ array with zero bulk field (bottom row). 
		The third column depicts the field configurations. The negativity of spin pairs joined by a solid (dashed) line
		in the bottom right panel is shown in the bottom left (central) panel.  $\Delta=1.2$ in all cases.}
	\label{fsm2}
\end{figure}

We now discuss the exact GS pair negativities $N_{ij}$ for a cyclic chain with next alternating fields (Fig.\ \ref{fsm2} top row)
and with zero bulk fields (central row), as well as for an open array with a zero
bulk field configuration (bottom row). The definite magnetization GS plateaus (see Fig.\ 4)
lead to $2Ns+1$ steps in $N_{ij}$, which coalesce exactly at the factorization point.

{\it Spin chain with next alternating fields.} The first-neighbor pairwise entanglement  (top left panel)
shows an asymmetric behavior, with the negativity of the 2-3 pair (a spin with zero field and a spin with applied field $h_2$) 
becoming maximum for large positive $M<Ns$, i.e.\ strong $h_1$ and weak $h_2$, and then decreasing as
$M$ decreases below $Ns/2$, i.e.\ as $h_2$ increases, since these spins become aligned with the field $h_2$. Second neighbor pair
entanglement for spins at even sites (top central panel), i.e.\ between spins with zero field, become  in contrast
appreciably entangled in the $M=0$ plateau, with negativity even increasing with increasing fields of
opposite sign. Due to field induced frustration these spins
are in entangled reduced pair  states with $\langle S_z^2 \rangle=\langle S_z^4 \rangle=0$ when $M=0$.
On the other hand, as $|M|$ increases $N_{24}$ decreases  as these spins become aligned with the stronger field.

{\it Spin chain with zero bulk fields.} This extremal factorizing field configuration corresponds to
the {\it minimum complexity} configuration for the spin chain as it requires just the application of
two nonzero fields of opposite sign. As seen in the central panels of Fig.\ \ref{fsm2}, the negativity
$N_{45}$ of two adjacent spins with no field (left) is maximum at the $M=1$ plateau but remains
entangled at the $M=0$ plateau, decreasing then for decreasing $M$ as the spin at site $j=5$ becomes
aligned with the closer field $h_2$ to contribute to the negative magnetization.
On the other hand, the pair negativity between spins at sites $i=4$ and $j=10$ (center), i.e.,
at two sites with zero field equidistant from those with fields (1 and 7), presents a significant field induced
entanglement at the $M=0$ plateau despite their large separation,
which is essentially similar in origin to the one previously discussed and which does not diminish with increasing
opposite fields.

{\it Open $3\times4$ array with zero bulk fields.} The bottom row of Fig.\ \ref{fsm2} depicts 
the negativities of the spin pairs schematically indicated on the right panel. The negativity of 
the first neighbor pair of the left panel (a bulk spin with no field and an edge spin with field $h_2$) 
presents an antisymmetric behavior similar to that depicted in the top left panel,  with pair negativity 
decreasing when M decreases,
while that of the second neighbor pair of the central panel (two edge spins with zero and $h_1$ fields) 
shows a more flattened behavior and decreases when $M$ increases, i.e. as the spins become aligned 
with the $h_1$ field.
That of the present two bulk spins (not shown) exhibits a symmetric behavior similar to that of the 
central panel of the second row.

\begin{figure}[t]
	\centering{\scalebox{.61}{\includegraphics{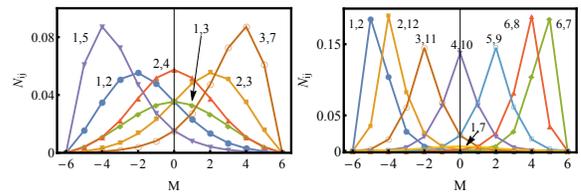}}}
	\vspace*{-0.0cm}
	\caption{Negativity at factorization ($h_1=-h_2=2h_s$)  in the definite magnetization projected GS's  as a function of
		$M$  for distinct two spin states (sites indicated), for an $N=12$ cyclic spin-$1/2$ chain with
		next alternating fields (left) and with zero bulk fields (right).}
	\label{fsm3}
\end{figure}

Fig.\ \ref{fsm3} depicts the negativity at factorization in the definite $M$ projected ground states, 
for a cyclic spin chain with next alternating fields and with zero bulk field (top and center chains 
in previous Fig. \ref{fsm2}). The GS structure at factorization
is more complex and leads to several distinct reduced pair states at this point,
whose negativities become
maximum at different $M$ values. For next alternating fields,
the maximum pair negativity is reached for two spins with the same field (1-5, 1-9, etc. or 3-7, 3-11, etc.) 
at finite $M$ of sign opposite to that of the corresponding field, while remaining pairs reach a lower maximum,
attained at $M=0$ for pairs with zero (2-4, 2-6, etc.) or opposite (1-3, 1-7, etc.) fields. 
In the zero bulk cyclic case we have just plotted the four most prominent pair negativities, which exhibit 
rather sharp maxima at different values of $M$. The maximum negativity is reached by pairs like 6-8 or 
6-7 at magnetization opposite to that of the applied field (at site 7), while distant zero field pairs 
symmetrically located from both fields (4-10) also reach a significant maximum at $M=0$. Selection of $M$ 
enables then to entangle different types of pairs.

\section{Appendix E. Counting spin and factorizing field configurations}
As discussed in the main body, in open chains of $N$ spins with first neighbor couplings,
after starting from an arbitrary  seed $\vartheta_0\neq 0, \pi$ at the first spin, there are
two possible alignment direction choices for each of the next spins according to Eq.\ (7) ($\theta_{i+1}=\vartheta_{k\pm 1}$
if $\theta_i=\vartheta_k$), leading to $2^{N-1}$ distinct FF configurations.
For instance, for constant exchange anisotropy $\Delta$ and coupling strength $J$, and $N=4$ spins $s$, 
we obtain  the eight FFs $\pm h_{\rm s}(1,-2,2,-1)$, $\pm h_{\rm s}(1,0,-2,1)$,
$\pm h_{\rm s}(1,-2,0,1)$ and
$\pm h_{\rm s}(1,0,0,-1)$, corresponding to product eigenstates
$|\vartheta_0\vartheta_{\pm 1}\vartheta_0\vartheta_{\pm 1}\rangle$,
$|\vartheta_0\vartheta_{\pm 1}\vartheta_{\pm 2}\vartheta_{\pm 1}\rangle$, 
$|\vartheta_0\vartheta_{\pm 1}\vartheta_0\vartheta_{\mp 1}\rangle$,
and $|\vartheta_0\vartheta_{\pm 1}\vartheta_{\pm 2},\vartheta_{\pm 3}\rangle$. In cyclic chains there are just 
$\binom{N}{N/2}$ configurations since the 1-$N$ coupling implies, for constant $\Delta$, $\theta_N=\vartheta_{\pm 1}$. 
Thus, for $N=4$ just the first six previous separable eigenstates
remain feasible, which lead  to FFs $\pm 2h_{\rm s}(1,-1,1,-1)$, $\pm 2h_{\rm s}(1,0,-1,0)$, and $\pm 2h_{\rm s}(0,-1,0,1)$.

For a spin-star geometry,  where a central spin  is coupled to $N-1$ noninteracting spins
(Refs.\ [55,48]), there are again $2^{N-1}$ FF configurations, according to the signs chosen in (7) 
for each coupling. A constant FF $h_{\rm s}$ at the $N-1$ spins is then feasible for constant $\Delta$, 
$J$ and $s$,  with a central field $-(N-1)h_{\rm s}$. A configuration with no field at the central spin 
is also possible if remaining fields satisfy the zero sum condition (9).

In the case of two-dimensional open rectangular arrays of $M\times N$ spins with first neighbor couplings 
and fixed exchange anisotropies $\Delta$ (for both vertical and horizontal couplings),  the determination 
of the number $L(M,N)$ of feasible
configurations is not as straightforward. For an open $2\times N$ array it is still easy to see that for 
a given seed angle $\vartheta_0\in(0,\pi)$, there are
\begin{equation}\tag{E1} L(2,N)=2\times 3^{N-1}\label{l2n}\end{equation} possible separable eigenstates and 
factorizing field configurations, since for any pair $(\vartheta_{k},\vartheta_{k+1})$ there are three possible continuations:
$(\vartheta_{k+1},\vartheta_{k+2})$ and  $(\vartheta_{k\pm 1},\vartheta_{k})$.

For an open $3\times N$ array a similar procedure leads to a system of first-order linear recurrences, 
from which we find a number of configurations given by
\begin{equation}\tag{E2}
L(3,N)=\alpha_+\lambda_+^N+\alpha_-\lambda_-^N\,,
\label{A2}
\end{equation}
where $\lambda_{\pm}=\frac{5\pm\sqrt{17}}{2}$, $\alpha_{\pm}=\frac{1\pm 3/\sqrt{17}}{2}$. This result yields, for instance, $L(3,3)=82$ configurations for a $3\times 3$ array.  The $4\times N$ and $5\times N$ cases can be similarly solved.

\begin{figure}[t]
	\vspace*{0.5cm}
	
	\centering{\scalebox{.6}{\includegraphics{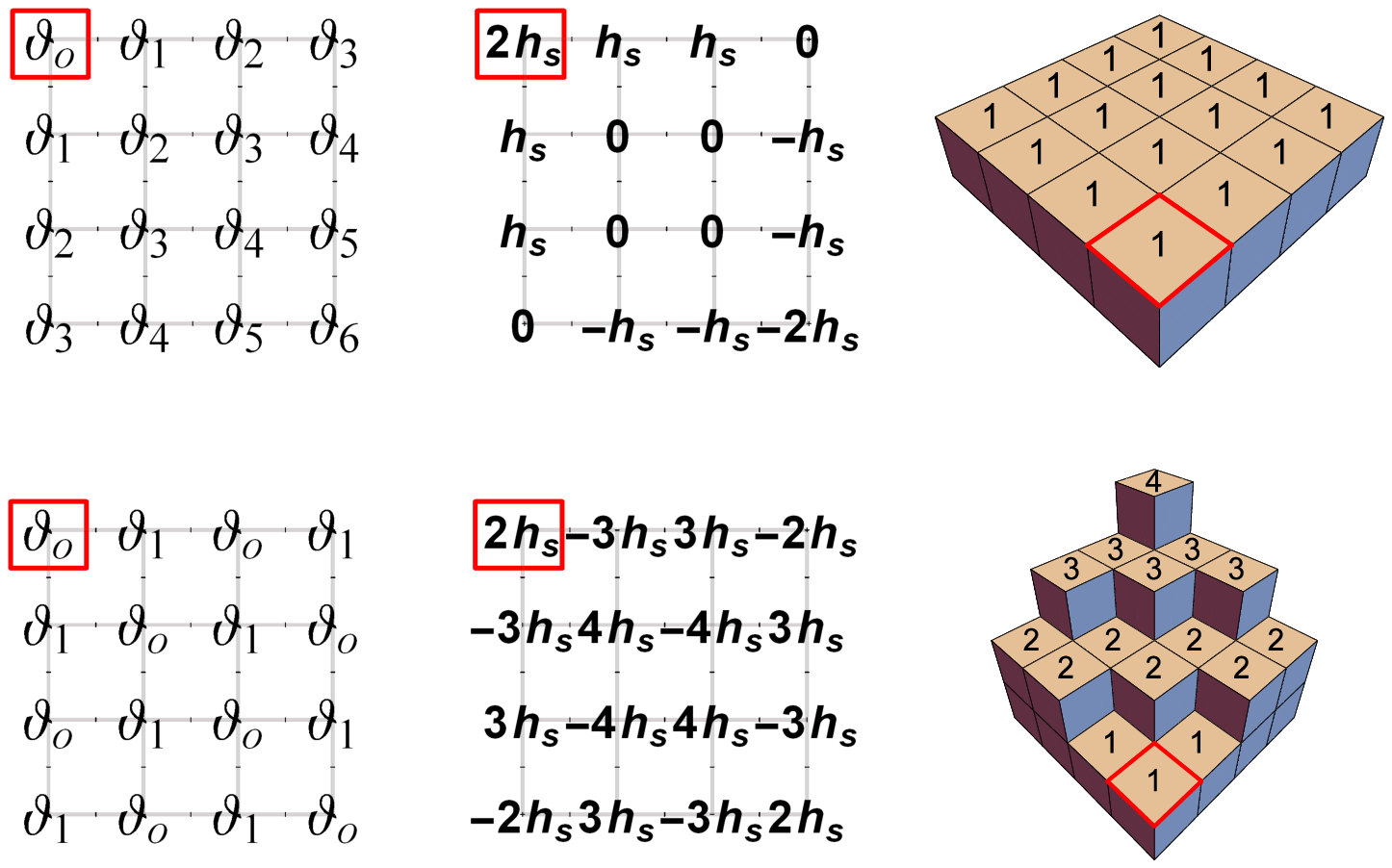}}}
	\vspace*{0.0cm}
	\caption{Schematic representation of the two extremal spin orientation angles and factorizing field configurations for a $4\times 4$  spin array: The solution with zero bulk factorizing fields (top) and the alternating field case (bottom). The third column depicts 
		the nondecreasing function $(i+j-k)/2$, with $i,j$ denoting the site and $k$ the steps from the initial seed of the orientation angle $\theta_{ij}=\vartheta_k$.}
	\label{fsm4}
\end{figure}

In the general case, given a factorized spin orientation configuration in an $M\times N$ array,  with a seed value $\vartheta_0$ at site $i=j=1$, the nondecreasing function defined as
\begin{equation}\tag{E3}
(i+ j-k)/2\,,
\label{E3}
\end{equation}
where $i,j$ indicates the site and $k$ is the number of steps in (7) from $\theta_{11}=\vartheta_0$ to $\theta_{ij}=\vartheta_k$, creates a one-to-one  correspondence between each factorized configuration and a two-dimensional terrace form array. The latter is composed of $M\times N$ integers that are nondecreasing both from left to right and top to bottom, such that two adjacent entries differ by at most $1$ (see right panel of Fig.\ \ref{fsm4}). 
Remarkably, the problem of counting the terrace forms spanned by Eq.\ (\ref{E3}) is equivalent to that of counting Miura-ori foldings 
\cite{GH.14} and to that of determining the number of ways to properly $3$-vertex-color an $M\times N$ grid graph with one vertex 
pre-colored \cite{M.15}. Although there are no known closed expressions such as (\ref{A2}) for a general $M\times N$ array, a 
recursive transfer matrix $A$ can be used to determine the number of spin and factorizing field configurations. By defining the matrices
$A(1)=(1)$ and
\begin{equation}\tag{E4}
A(M+1)= \left(
\begin{array}{cc}
A(M) & A(M)^T \\
0 & A(M) \\
\end{array}
\right)\,,
\end{equation}
with $B(M)=A(M)+A(M)^T$, then the total number of spin orientation angles configurations (and hence of FF configurations) for a given seed is given by
\begin{equation}\tag{E5}
L(M,N)=\sum_{i,j}(B^{N-1}(M))_{ij}\,,\label{Eg3}
\end{equation}
i.e., the sum of all the entries of $B^{N-1}(M)$. For $M=2$ and $3$ Eq.\ (\ref{Eg3}) leads to previous results (\ref{l2n})--(\ref{A2}). 
 For large $M$ and $N$ the number of total configurations grows exponentially with the dimension \cite{GH.14,Li.67}.

\section{Appendix F. Physical Realization}
As mentioned in the introduction, there are presently various promising schemes for simulating $XXZ$ arrays
with tunable couplings and fields. In first place, cold atoms in optical potentials provide a convenient platform.
The effective Hamiltonian of strongly interacting two-component atoms confined in a one-dimensional trapping potential 
can be mapped onto the spin $1/2$ $XXZ$ model after denoting the Bose-gas components as spin down or up, 
with the effective fields $h^i$  depending on the inhomogeneous applied field $B$  and the first neighbor 
coupling strengths by the contact interaction between the atoms (see for instance Ref.\ [28]). 
The effective anisotropy $\Delta$ can be controlled through the parameters of the trapping potential. 
The recent proposal of Ref.\ [29] is based on laser trapped Rydberg atoms in circular states, 
i.e.\ states with maximum angular momentum, which exhibit very long lifetimes. The up and down 
spin states are circular states with different principal quantum number and the $XXZ$ coupling 
emerges from the large dipole-dipole coupling, with $J$ depending on the interatomic distance and $J_z$  
tunable through the static electric field.
Application of a further classical field with appropriate polarization and frequency leads to effective 
frequency dependent constant fields along the $z$ and $x$ directions in the final $XXZ$ Hamiltonian. 
While these fields are uniform except for border corrections  if the applied field is uniform, 
application of a nonuniform field would lead to a nonuniform effective field. We remark here that 
among the several possible FF configurations discussed in Appendix D and in the main body, 
some of them are of  low complexity, like the zero-bulk field configuration requiring just 
opposite edge effective fields, i.e.\ adequate border corrections.

Other possible realizations include those based on trapped ions [30,43-46] and quantum dots [48]. 
In [48] a scheme based on two or more electron spins in a linear geometry, which provides the ``bus'', 
plus  additional spins which generate the qubits, lead to an effective $XXZ$ coupling between the qubits. 
Such interaction emerges from an isotropic-like Heisenberg coupling between the electron spins of the bus 
and qubits via second order perturbation. The resulting Hamiltonian contains an effective field acting 
on the qubits, which depends on the local tunable magnetic field applied to the qubit spins, and on a 
 perturbative correction leading to an alternating longitudinal field (determined by the bus-qubit coupling). 
 The ensuing effective coupling affects in principle any two qubit spins. Nonetheless,  
 a critical regime is also possible in which the magnetic field on the (even-size) 
 bus is tuned to be near a ground state level crossing. In such a case, a first order $XXZ$  
 coupling between each qubit and the bus emerges, whose strength is of similar magnitude 
 as the original bare exchange coupling, with the bus represented by an effective qubit 
 based on the two crossing bus states. The ensuing Hamiltonian corresponds at first order 
 to a spin-star like architecture (a bus qubit coupled to $N$ noninteracting qubits [55]), 
 with the field on the qubits again tunable since it depends on the original applied field 
 and a perturbative correction along $z$.
As  mentioned in the previous section, such geometry allows for low complexity FF configurations, 
such as that with a constant field on the qubits and an opposite field on the bus qubit, 
feasible for constant bus-qubit coupling and any number $N\geq 1$ of qubits. Nevertheless, 
the present factorization will also arise even if the qubit-bus $XXZ$ couplings 
are nonuniform as long as $|\Delta_{ij}|\geq 1$, with the factorizing fields obtained from Eq.\ (8). 

Another possible realization of an XXZ Hamiltonian with tunable interactions and fields 
can be  achieved using superconducting charge qubits (SCQ). For instance, in the scheme 
discussed in Ref.\ [41], the qubits are realized by superconducting islands coupled 
to a ring by two symmetric Josephson junctions, so that the states $|0\rangle$ 
and $|1\rangle$  correspond to the two charge states near the charging degeneracy 
point (in the charging regime the extra Cooper-pairs number $n$ in the 
box is a good quantum number, while near the charging energy degeneracy point, $n=0,1$). 
If a control gate voltage is applied to each SCQ box through a capacitance
 and an external magnetic flux is used to modulate the Josephson coupling energy, 
 then the local field parameters can be tuned.
Finally, qubit-qubit interactions are achieved by coupling the SCQ's with a 
superconducting quantum interference device (SQUID) pierced by a magnetic
 flux which can be used to control the Josephson coupling.
The effective interaction is of the $XXZ$-type with tunable strength.


\begin{thebibliography}{999}
\bibitem{note} As $(\theta,\phi+\pi)$ is equivalent to $(-\theta,\phi)$, it is unnecessary to consider solutions
with $\phi_{ij}=\pi$ (which  also satisfy Eqs.\ (\ref{1}) and (\ref{4})) if negative values of $\theta$ are allowed.
\bibitem{Szm} $\langle S_i^z\rangle_M$ is the average at the projected states (\ref{PM2})--(\ref{redp}).
\end{thebibliography}
\end{document}